# Nonlinear Photoluminescence Imaging of Isotropic and Liquid Crystalline Dispersions of Graphene Oxide

Bohdan Senyuk,[†,‡] Natnael Behabtu,[‡] Benjamin G. Pacheco,[†] Taewoo Lee,[†] Gabriel Ceriotti,[§] James M. Tour,[§] Matteo Pasquali,[‡,§,]* and Ivan I. Smalyukh,[†,⊥,∥,]*

[†]Department of Physics and Liquid Crystal Materials Research Center, University of Colorado, Boulder, Colorado 80309, United States

[‡]Department of Chemical and Biomolecular Engineering, [§]Department of Chemistry, R.E. Smalley Institute for Nanoscale Science & Technology, Rice University, Houston, Texas 77005, United States

[⊥]Department of Electrical, Computer, and Energy Engineering and Materials Science Engineering Program, University of Colorado, Boulder, Colorado 80309, United States

[∥]Renewable and Sustainable Energy Institute, National Renewable Energy Laboratory and University of Colorado, Boulder, Colorado 80309, United States

*Address correspondence to mp@rice.edu (M.P.) and ivan.smalyukh@colorado.edu (I.I.S.).

**Abstract**

We report a visible-range nonlinear photoluminescence (PL) from graphene oxide (GO) flakes excited by near-infrared femtosecond laser light. PL intensity has nonlinear dependence on the laser power, implying a multiphoton excitation process, and also strongly depends on a linear polarization orientation of excitation light, being at maximum when it is parallel to flakes. We show that PL can be used for a fully three-dimensional label-free imaging of isotropic, nematic, and lamellar liquid crystalline dispersions of GO flakes in water. This nonlinear PL is of interest for applications in direct label-free imaging of composite materials and study of orientational ordering in mesomorphic phases formed by these flakes, as well as in biomedical and sensing applications utilizing GO.

# INTRODUCTION

Unique mechanical,[1] thermal,[2] optical,[3,4] and electronic[3,5] properties of graphene make it a subject of significant scientific interest and a material of choice in modern photonics, optoelectronics, and nanotechnology.[3,4] The variety of graphene applications ranges from transparent electrodes in display technology[4,6] to organic photovoltaics.[7,8] Graphene oxide (GO) is also attractive to researchers and engineers because of its easy processing and potential for the large-scale production of graphene-based materials,[4] as it retains most of the mechanical and electronic properties of graphene (possibly after reductive postprocessing).[9] Owing to its properties, GO has been used to spin fibers[10,11] and make free-standing paper[12] as well as for cellular imaging[13] and biodetection.[14] Unlike a zero band gap graphene,[9] GO in aqueous dispersions or solid state readily shows a broad-band photoluminescence (PL) in the visible[13,15-17] and near-infrared (NIR)[13,15] range excited by near- or medium-UV, thought to be a result of $\pi$-$\pi$* interband transitions.[9,13,15-17] Galande et al.[18] reported a quasi-molecular visible fluorescence from GO similar to polycyclic aromatic fluorophores. However, a nonlinear PL at wavelengths shorter than excitation, which was found in graphene,[19,20] has not been so far reported from GO flakes (GOFs). Finding the nonlinear optical response from GO would be interesting from the basic physics point of view and exciting for new applications, which prompted us to experimentally address this problem.

In addition to its exciting optical properties,[3,9] GO is water-soluble, which makes it very interesting for soft matter physics and useful for nanotechnology. Recently, it was shown that, similar to graphene dissolved in strong acids,[21] GO flakes dispersed in water form various liquid crystalline phases,[11,22-24] fluids with an orientational order of constituents.[25] The interest in liquid crystals (LCs) made of aqueous plate-like GOFs is two-fold: first, they can be a precursor for developing the graphene-based nanostructures and devices, and second, LC phases formed of plate-like colloidal particles are important for fundamental studies.[25,26] Until recently, LC structures and orientational order have been studied with conventional polarizing optical microscopy (PM).[25] Fluorescent confocal polarizing microscopy[27] opened the possibility of a three-dimensional mapping of long-range orientational order in dye-doped LCs and other soft matter systems. However, label-free three-dimensional imaging of soft matter materials with increased spatial resolution can be achieved with nonlinear optical microscopy techniques[28-30] via multiphoton excited fluorescence of their constituents. The nonlinear multiphoton imaging has many unique advantages compared to PM such as a superior spatial resolution due to a highly localized excitation, an enhanced penetration depth, and reduced photodamage of the specimen due to NIR excitation.[31]

In this article, we report a strong visible PL from aqueous and dried drop-casted GOFs illuminated by NIR femtosecond laser light. The intensity of PL has nonlinear dependence on the input excitation power and strongly depends on the polarization of excitation light, similar to a nonlinear multiphoton excitation fluorescence from organic molecules and fluorophores.[29-31] We use this nonlinear PL for fully three-dimensional label-free imaging of structures of isotropic and orientational order of liquid crystalline phases in LC samples formed by aqueous GOFs. Furthermore, the nonlinear orientation-sensitive PL from GO can be also used for imaging in other soft matter systems, in biomedical applications,[13] as drug delivery vehicles,[32,33] and most importantly in graphene-based composite materials and structures,[3,34] as well as can expand applications of GO.

## RESULTS AND DISCUSSION

### Characterization of Photoluminescence from GO Flakes.

Optical absorbance spectra of aqueous vigorously sonicated GO flakes (see Methods) show characteristic peak at ~240 nm (Figure 1a) likely caused by π-plasmon absorption of carbon.[13,17,18] Samples transmit light mostly in the long-wavelength part of the visible spectral range (Figure 1a) and have a light brown color appearance; more concentrated and thick samples become opaque. Illumination of both aqueous and dried drop-casted (solid) GO flakes with a pulsed femtosecond laser beam in the range of ~800-1000 nm results in the corresponding broad-band PL in the visible spectral range (Figure 1b). The functional form of the spectral line shape remains unchanged as the excitation power, concentration, and thickness of samples are varied, although it appears to be somewhat blue-shifted for longer excitation wavelengths (Figure 1b). For most of the nonlinear imaging results reported here, we chose the excitation wavelength at 850 nm (≈1.46 eV). The resulting PL shows highly nonlinear dependence on the intensity of excitation light. The slope of a log-log plot of PL intensity versus laser power (Figure 1c) changes from ~2 at lower power to ~3 at higher intensity of excitation. This dependence implies that the observed PL is driven by a quasi-molecular[13,18] multiphoton absorption process, which is two-photon at lower and three-photon at higher excitation powers.

Importantly, the GOFs' nonlinear PL is sensitive to the polarization of excitation light, as discussed below in the context of three-dimensional imaging of long-range orientational order of LCs formed by aqueous GO flakes.

Figure 2a,b shows PL images of as-obtained and tip-sonicated GO flakes, which allow for measuring approximately their size distributions shown in Figure 2c. The different number of layers in GO flakes causes the PL intensity difference between flakes (Figure 2a): intensity increases as the number of layers increases.[19] The imaging of small flakes is limited by optical resolution of several hundreds of nanometers.[29,30] The distributions of PL images of flakes are skewed toward smaller flakes and can be fitted with a log-normal function (see Methods), which yields an average equivalent diameter ($D_{GO}$), its standard deviation ($\sigma$), and polydispersity ($p = \sigma/D_{GO}$). The shape of flakes can be characterized by circularity ($C_r$). The original flakes (Figure 2a,c) are large ($D_{GO}$ = 2.77±0.3 μm) and have a very irregular, often anisotropic, shape ($C_r \sim 0.8$). They are very polydisperse ($p = 0.67$), with flake sizes ranging from hundreds of nanometers to ~30 μm. However, an additional, comparatively short,[35] tip sonication of the original GOFs results in more narrowly disperse ($p = 0.33$) flakes (Figure 2b,c) of smaller size ($D_{GO}$ = 0.48±0.3 μm) and round shape (see inset in Figure 2c), which were used in the LC samples for nonlinear imaging.

### Nonlinear Imaging of Aqueous GO Liquid Crystals.

Above a certain critical concentration (dependent on the actual size distribution[22]), aqueous GO flakes form a discotic nematic LC phase.[11,22-24] In our experiments, the aqueous dispersion of vigorously sonicated GOFs exhibits a nematic phase above the concentration of ~0.25 wt %. The director ***n***, describing the average local orientation of LC building blocks,[25] is normal to the plane of disk-like flakes[22] (Figure 3) and parallel to their optic axis; the optical anisotropy of the GO LC is $\Delta n = n_\parallel - n_\perp < 0$, where $n_\parallel$ and $n_\perp$ are refractive indices measured for light polarized parallel and perpendicular to ***n***.[22] Figure 3a,b shows a PM texture of a flow-aligned GO nematic LC at 0.7 wt % confined between two untreated glass slides: ***n*** is tangential to the confining slides and oriented at 45° between crossed polarizers "P" and "A". A retardation plate with a "slow" axis ***γ*** additionally verifies the orientation of ***n***: blue color in the texture corresponds to orthogonal ***n*** and ***γ***, whereas yellow color indicates parallel ***n*** and ***γ***.[22]

The intensity of PL from the aligned GO nematic strongly depends on the angle *β*

between the plane of the flake and the polarization $e$ of excitation light (Figure 3f). The PL intensity changes from minimum at $\beta = 90°$ ($e\|n$) to maximum when $e \perp n$ roughly as $I_{PL} \propto \cos^4\beta$, which is another indication of a twophoton[29-31] quasi-molecular[18] excitation process (a function $I_{PL} \propto \cos^2\beta$ expected for a linear process gives a much worse fit). Nonlinear PL images taken for two orthogonal $e$ are shown in Figure 3c,d: the PL intensity matches perfectly with the orientation of $n$ deduced from PM textures (Figure 3a,b). This demonstrates the feasibility of orientation-sensitive nonlinear imaging based on the polarizing PL from GO flakes. The nonlinear optical nature[31] of the observed PL allows for submicrometer axial resolution[29,30] and the fully three-dimensional imaging of GO samples (Figure 4), as demonstrated using vertical cross-sectional PL images of a thick (≈24 μm) LC cell (Figure 4h,i). The detailed director field $n(r)$ of the Schlieren texture in the GO nematic (at 0.5 wt % of GO) confined between untreated substrates (Figure 4g) and reconstructed from PL images by changing the orientation of $e$ (Figure 4b, c,e,f) is consistent with $n(r)$ deduced from PM textures (Figure 4a,d). It shows two half-integer $k = +1/2$ line defects, where $k$ is the strength of disclination defined as a number of times $n$ rotates by $2\pi$ when the defect core is circumnavigated once.[25] The stability of two $k = +1/2$ as compared to one $k = +1$ line defect is natural because their combined elastic energy is twice smaller since the elastic energy of a line defect[25] is $W \propto k^2$. This behavior is also consistent with the fact that $k = \pm 1$ disclination in the GO LC does not escape into the third dimension.[25] Figure 4h,i reveals $n(r)$ in the vertical direction across the LC cell, which cannot be directly visualized by the PM.

Recently, Dan et al.[22] showed that foreign inclusions of sufficient size distort $n$ when introduced into GO nematic LCs. We therefore probe $n(r)$ around solid particles with PL imaging as another example of its imaging capabilities. Figure 5 shows PM and PL images of $n(r)$ distortions around a spherical glass particle of radius $R = 10$ μm immersed into the aqueous GO nematic dispersion at 0.7 wt %. Both PM and PL images verify a homeotropic (normal to the surface) anchoring of $n$ at the sphere's surface and show weak director distortions (green lines in Figure 5g) of a quadrupolar symmetry[22,36] around the solid microsphere. No singular disclination loop or point defect is observed, suggesting a weak surface anchoring[37] for GOFs estimated to be $W_a \sim K/R \sim 10^{-7}$-$10^{-5}$ J m$^{-2}$, where $K \sim 10^{-12}$-$10^{-10}$ N is an average elastic constant[22] dependent on the size distribution of the flakes.[25]

Highly concentrated aqueous GO dispersions can form lamellar-like phases.[11] Figure 6

shows fan-shaped textures of highly concentrated (1.2 wt %) aqueous dispersion of GOFs confined between untreated glass substrates. Textures show misaligned stripes and domains with sharp edges and boundaries corresponding to strong folding[38] or buckling (Figure 6i) of a layered structure (Figure 6j) stabilized by electrostatic interactions[11] between negatively charged GO flakes and by hydrogen bonding[39] (Figure 6k). PL images (Figure 6f,g) of fan-shaped textures are consistent with PM observations (Figure 6e) but provide more detailed information about $n(r)$, in particular in-plane sections across the sample. PL images taken at orthogonal polarizations of excitation (Figure 6b,c,f,g) are complementary to each other and yield unambiguous direct information about the director field (Figure 6d,h). They demonstrate that the nonlinear PL is suitable for direct imaging of orientational order not only in nematic but also in lamellar LC phases formed by GO flakes.

## CONCLUSIONS

We have observed the visible-range PL from aqueous and solid GO flakes excited by near-infrared femtosecond laser light. The PL intensity shows nonlinear dependence on the input excitation power, implying a multiphoton excitation process: two-photon at lower and three-photon at higher excitation power. Also, the PL intensity depends strongly on the polarization of excitation light; it is the largest when the polarization is parallel to the plane of GO flakes. Sensitivity to the direction of excitation linear polarization makes it suitable for imaging not only positional but also orientational order in different soft matter systems. To show this ability, we have used the nonlinear PL for fully three-dimensional label-free imaging of structures of isotropic and orientational order of nematic and lamellar liquid crystalline phases in LC samples formed by aqueous GO flakes. Additionally, using PL imaging, we have demonstrated that the size and polydispersity of GO flakes can be significantly altered by vigorous tip sonication. Although the underlying physical origins of the observed optical fluorescence and photoluminescence effects in GO are still not well understood,[9,15,18] the nonlinear orientation-sensitive PL from GO is attractive for a wide range of applications and can be also used for cellular imaging in biology,[13] probing drug delivery using GO flakes,[32,33] and for orientation-sensitive imaging in graphene-based composite materials and structures[3,34] needed for energy conversion applications.[7,8] Moreover, this highly efficient nonlinear PL can potentially lead to novel applications of GO utilizing its photoluminescent properties.

## METHODS

**Aqueous Dispersions of Graphene Oxide Flakes and Materials.** The original, mostly single-layer, improved GO flakes were synthesized by a large-scaled version of methods described elsewhere[40-42] and delivered as a dispersion in deionized (DI) water at 2.5 g/L. The synthesis and purification procedure was as follows: 3 g of xGnP 5 μm graphite nanoplatelets (XG Sciences, Inc.) and 200 mL of 9:1 $H_2SO_4/H_3PO_4$ solution were added to a 1 L Pyrex beaker. The mixture was set to stir continuously using an IKA Werke RW 16 mechanical stirrer. Slowly, 9 g of $KMnO_4$ was stirred into the mixture, which turned a deep dark green color. Precautions should be taken to not exceed ~5 wt % of $KMnO_4$ per addition and make sure the change in color from green to purple is complete before introducing more oxidant as solutions with more than 7 wt % of $KMnO_4$ added to $H_2SO_4$ can explode upon heating.[43] The mouth of the beaker was covered as much as possible using aluminum foil, and the reaction was heated to 40°C using a resistance-heated water bath. After 6 h, the mixture had thickened and turned from dark green to purple-pink. At this point, 9 g more of $KMnO_4$ was added and allowed to react for another 6 h. When the reaction was completed, the resulting slurry was quenched over 200 mL of ice-water containing 5mL of 30% $H_2O_2$. The resulting bright yellow GO suspension was purified by a series of centrifugations and resuspensions in clean solvent. All centrifugations were performed at 4000 rpm for 90 min, and all resuspensions were performed by shaking the centrifuged GO for ~4 h at 200 rpm on a platform shaker using 200 mL of solvent. The product was centrifuged once as-is and centrifuged washed once in DI $H_2O$, once in 30% HCl, and twice more in DI $H_2O$. To determine the concentration of the final solution, a 10 mL aliquot was filter-washed with 200 mL of three different solvents: methanol, acetone, and diethyl ether. The final material was vacuum-dried (4 Torr) at room temperature overnight and weighed (25 mg).

To obtain more monodisperse flakes of smaller size, the original GOF dispersion was additionally tip-sonicated for 2 h at ~35 W of ultrasonic power using a Branson 250 sonifier (VWR Scientific) operating at 20 kHz and equipped with a microtip of diameter 4.8 mm. The concentration of GOFs in LC samples was controlled by extracting water from the dispersion via centrifugation and evaporation. First, the dispersion was centrifuged for ~10 min at 12000 rpm in a Sorvall Legend 14 centrifuge (Thermo Scientific), and excess water was removed from the top of a vial by a calibrated pipet. To reach higher concentrations (>1 wt %), water was evaporated from the dispersion in the graduated vials.

The GOF dispersions were filled into the capillary or confined between two glass substrates. Fused silica rectangle (0.2×2 mm) capillaries (VitroCom) were used for optical characterization of the GOF dispersions. Clean untreated glass substrates were used to prepare GOF samples for microscopy observations, and 0.15 mm thick coverslips were used in the samples for multiphoton imaging. Stripes of Mylar film (DuPont Teijin) and glass spacers were used to set the thickness of a gap between confining substrates in a range of 10-30 μm. Dispersions of GOFs were additionally sonicated in a Cole-Parmer 8891 ultrasonic bath for ∼5 min before filling into the cell. Samples were sealed with an epoxy glue and nail polish on the top to prevent water evaporation. To measure the size distribution of GOFs, highly diluted (<0.05 wt %) aqueous dispersions were drop-cast on the clean glass substrate and baked at 100°C for ∼10 min to evaporate water. Spherical glass particles (Duke Scientific, Corp.) of a diameter 20 μm were used as foreign inclusions in the GOF LC samples. Microparticles in the form of dry powder were mixed directly into GO dispersions and sonicated in the ultrasonic bath for ∼15-30 min before filling into the cell to break apart pre-existing colloidal aggregates.

**Optical Microscopy Characterization and Nonlinear Imaging.** Optical properties of GOFs such as absorbance and transmission were measured using a Hewlett-Packard HP 8452A diode array UV/vis spectrophotometer. A spectrometer USB2000 (OceanOptics) mounted onto an inverted microscope IX-81 (Olympus) was used to detect PL spectra from GOFs. The average PL intensity was measured when exciting a constant area (115 × 115 μm) of the GO sample by scanning a tightly focused (area is ∼1 μm$^2$) laser beam. The average excitation laser power delivered to the sample was in the range of ∼0.5-10 mW. Polarizing microscopy observations of GOF dispersions in visible light were performed using IX-81 equipped with crossed polarizers, a 530 nm full-wave retardation plate, and a CCD camera (Flea, PointGrey).

Nonlinear imaging was carried out at room temperature using a multimodal nonlinear optical microscopy setup[29,30] coupled to the microscope IX-81. A tunable (680-1080 nm) Ti:sapphire oscillator (140 fs, 80 MHz, Chameleon Ultra II, Coherent) was used as an excitation source. The polarized excitation beam was focused into the sample using Olympus high numerical aperture (NA) oil objectives 100×/NA = 1.4 or 60×/NA = 1.42. The lateral position of the excitation beam in the plane of the sample was controlled with the Galvano-mirror scanning unit (Fluoview FV300, Olympus). The polarization of excitation was varied using a half-wave retardation plate mounted immediately before the objective. The excitation of GOFs was performed at 850 nm, and the unpolarized PL light in a range of 400-700 nm was detected in a backward mode with a photomultiplier tube H5784-20 (Hamamatsu) at a scanning speed of ~2 μs/pixel. To prevent the photo- and thermal damage, the average laser power in the sample was <1 mW for imaging. Data acquisition and image (512 × 512 pixels) reconstruction were performed with Olympus Fluoview software, and ImageJ software was used for data processing and analysis.

Using ImageJ built-in plugins, an equivalent diameter $D_{GO}$ of GO flakes was calculated as a diameter of a circle of the same measured area, and the circularity was calculated as $C_r = 4\pi(a/P^2)$, where $a$ and $P$ are an area and perimeter of the flakes, respectively. Size distribution was fitted with a log-normal function $y = B \exp[-\ln(x/x_c)^2/(2w^2)]$, where $\ln(x_c)$ is the mean, $w$ is the variance, and $B$ is a fitting constant.

**Acknowledgment.** This work was supported by NSF Grants DMR-0645461, DMR-0820579, and DMR-0847782, AFOSR Grants FA9550-09-1-0590 and FA9550-09-1-0581, and Welch Foundation Grant C-1668. We also acknowledge support from ICAM Branches Cost Sharing Fund, the AFRL through University Technology Corporation (09-S568-064-01-C1), and the Office of Naval Research Graphene MURI Program (00006766, N00014-09-1-1066). We thank B. Dan, Q. Liu, J. Evans, C. Twombly, R. Trivedi, A. Martinez, P. Ackerman, and M. Pandey for useful discussions.

# Figures

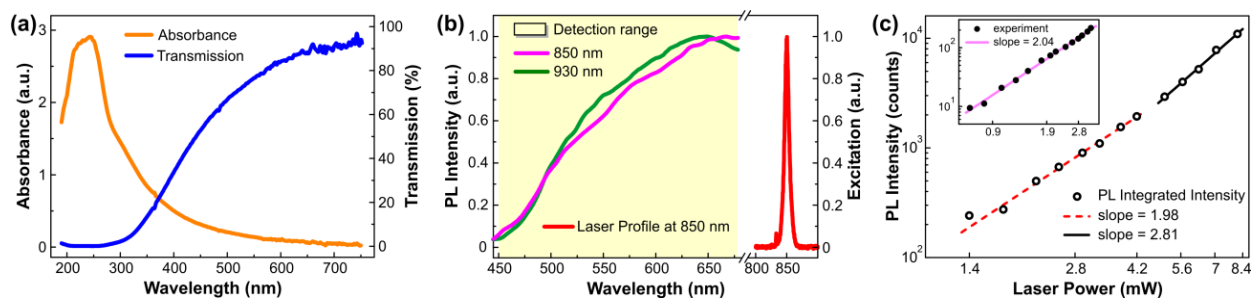

**Figure 1.** Nonlinear photoluminescence from aqueous GO flakes (0.25 wt %): (a) absorbance and transmission spectra; (b) detected normalized PL spectra for different excitation wavelengths at 2 mW and lasing spectrum at 850 nm. The yellow background shows a detection range. (c) PL intensity as a function of laser power. Quadratic dependence (slope≈2) implies a two-photon excitation process. Inset shows the PL intensity measured at low laser powers with a photomultiplier tube.

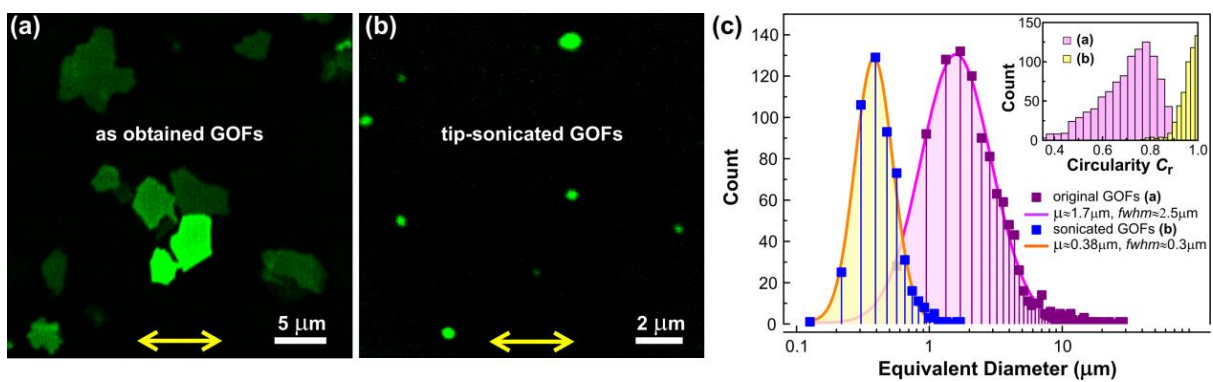

**Figure 2.** Size and shape distribution of GOFs spin-coated on glass substrates as measured using nonlinear microscopy: PL images of original as-obtained (850 nm at ∼200 μW) (a) and tip-sonicated (850 nm at ∼350 μW) (b) GOFs. Yellow arrows show the polarization of excitation light. (c) Size distribution of GOFs before and after tip sonication. Solid lines correspond to the fit of experimental data with a log-normal distribution with the mode $\mu$ and full width at half-maximum (fwhm). The inset shows the shape distribution of flakes.

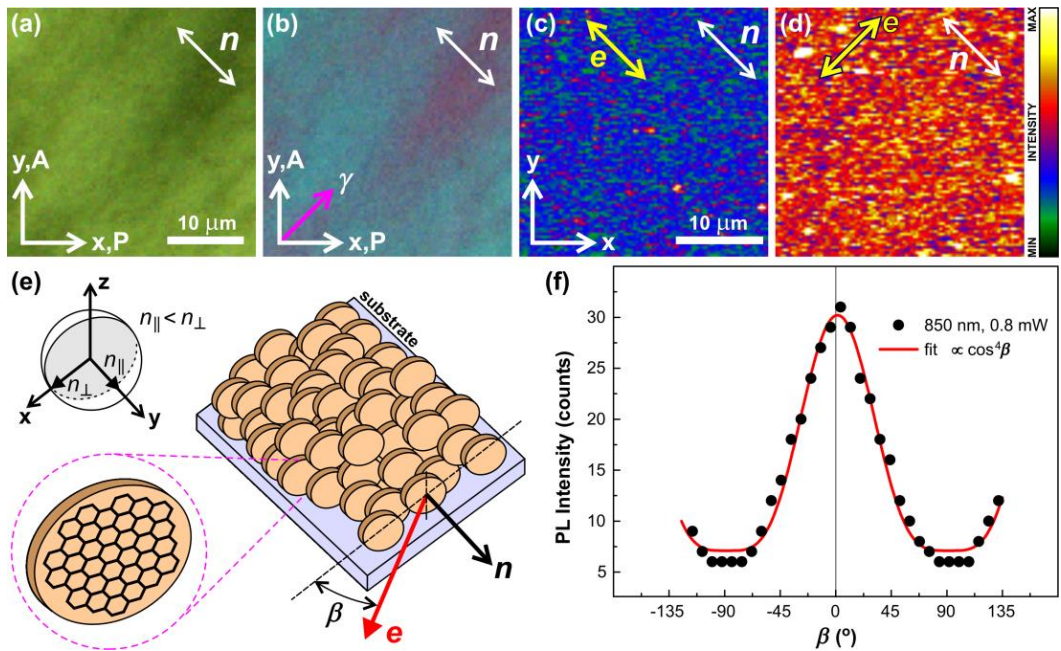

**Figure 3.** Orientation sensitive PL of aqueous GO flakes at 0.7 wt % in a nematic phase flow-aligned between two glass substrates ($d \approx 10$ μm): PM textures (a) between crossed polarizers and (b) with a retardation plate with a slow axis $\gamma$. Crossed polarizers are marked "P" and "A". (c,d) PL images of the area shown in (a,b) for two orthogonal polarizations (yellow arrows) of excitation light. The color-coded bar indicates the intensity of PL. (e) Schematic of GO flakes (brownish discs) aligned in a nematic phase along the director $n$ and illuminated by excitation light with a linear polarization $e$, and the indicatrix of their refractive indices. (f) Dependence of PL intensity on the angle $\beta$ between $e$ and the plane of flakes.

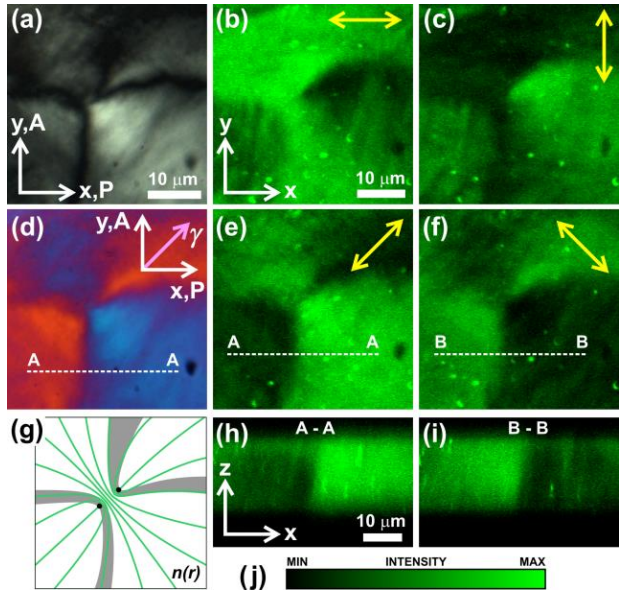

**Figure 4.** PM and PL textures of aqueous GOFs in a nematic phase at 0.5 wt % between untreated glass substrates in a cell of $d \approx 24$ μm: nematic texture between (a) crossed polarizers and (d) with a retardation plate. (b,c,e,f) In-plane ($xy$) PL images of area shown in (a,d) for different polarizations of the excitation light (yellow arrows) with average power of 0.8 mW at 850 nm. (g) Schematic of the corresponding $n(r)$ (green lines). (h,i) Cross-section ($zx$) PL images along the white dashed lines, respectively, in (d f). Gray brushes indicate areas of light extinction between crossed polarizers as in (a). Black filled circles show $k = +1/2$ disclination lines. (j) Color-coded bar shows the PL intensity scale.

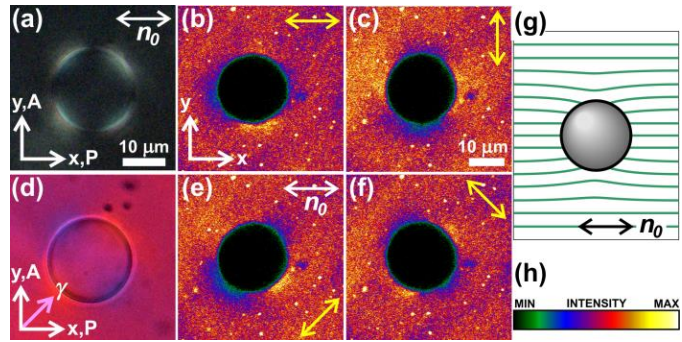

**Figure 5.** Textures of nematic GOFs at 0.7 wt % around spherical inclusions: PM texture of quadrupolar distortions (a) between crossed polarizers and (d) with retardation plate. (b,c,e,f) PL images of director distortions shown in (a,d) for different $e$ (yellow arrows). (g) Schematic of the corresponding $n(r)$ shown by green lines; $n_0$ is a far-field director. (h) Color-coded bar shows the PL intensity scale.

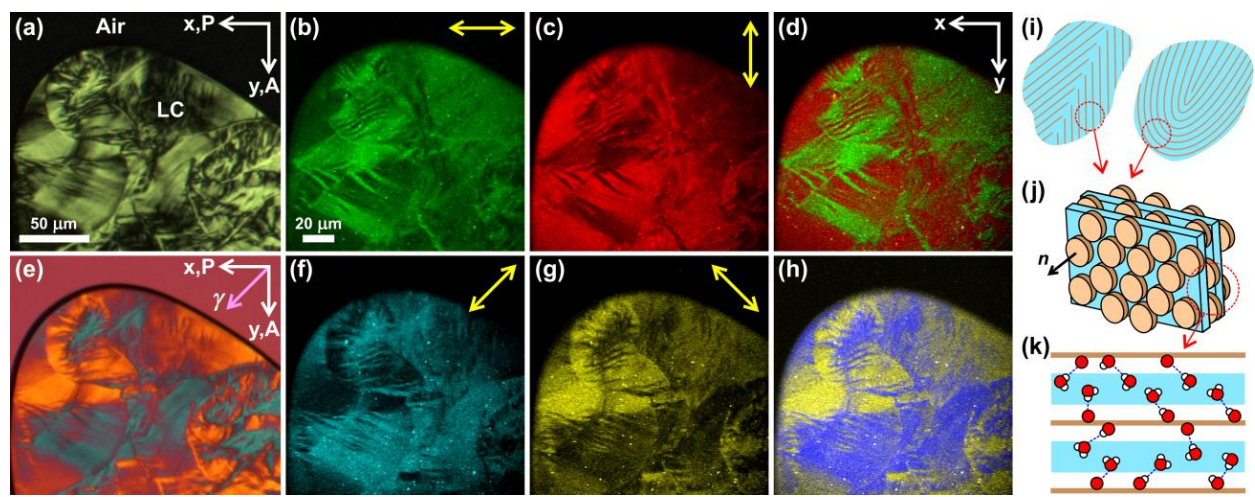

**Figure 6.** Fan-shaped textures of GOFs in a LC dispersion at higher concentration of 1.2 wt % confined between untreated glass substrates of a cell with $d \approx 16$ μm: PM textures between (a) crossed polarizers and (e) with a retardation plate. (b,c,f,g) PL images of LC texture shown in (a,e) for different $e$ shown by yellow arrows. (d,h) Overlaid PL images shown in (b,c) and (f,g), respectively. (i) Schematic of folding and buckling of GO layers (brown lines); blue color shows water. (j) Schematic of electrostatically stabilized GO layers. (k) Schematic showing hydrogen bonding between GOFs and water: white and red spheres show, respectively, hydrogen and oxygen atoms, and blue dashed lines show hydrogen bonds.